\documentclass[preprint,amssymb,nofootinbib]{revtex4}
\usepackage{fontenc}
\usepackage[latin1]{inputenc}
\usepackage{graphicx}
\usepackage{amsmath,amscd}
\usepackage{diagrams}

\begin{document}

\title{Finite-sites corrections to the Casimir energy on a periodic lattice}
\author{Michael Pawellek}
\affiliation{Department of Theoretical Physics, \\
 Royal Institute of Technology (KTH)- AlbaNova University Center, \\
 Roslagstullbacken 21, 106 91 Stockholm, Sweden}
\email{pawellek@kth.se}

\begin{abstract}
We show that the vacuum ground state energy for massive scalars on a 1-dim \(L\)-sites periodic lattice can be interpreted as
the thermodynamic free energy of particles at temperature \(1/L\) governed by the Arutyunov-Frolov mirror Hamiltonian.
Although the obligatory sum over zero-point-frequencies is finite on the lattice, a renormalization prescription is necessary in 
order to obtain a physical sensible result for the lattice Casimir energy. The coefficients of every term
in the large \(L\) expansion of the lattice Casimir energy are provided in terms of modified Bessel functions. 
\end{abstract}

\maketitle

\section{Introduction}

There are at least two main motivations for introducing a lattice structure in quantum models. In solid state physics one
introduce lattice models as the Heisenberg spin chain or the Ising model to study the properties of matter built out of a granular
atomic lattice.
In the realm of particle physics lattice models based on continuum quantum field theories such as gauge theories, 
especially QCD, serve as an UV cutoff and they enable to study non-perturbative effects.

The introduction of a periodic lattice with a finite number of \(L\) sites will affect the physical observables such as masses of
particles or the vacuum ground state. Although the literature refers to influences of finite lattices as 
finite-size effects (see e.g. \cite{Alba}), one should be aware that compared to the original QFT on continuous flat Minkowski space 
two steps of idealization are involved. 

The first one is {\it compactification} of space on a circle (or hypertorus in the case of several spatial dimensions) of size \(R\). The corresponding 
{\it finite-size} effects on masses of particles were initially studied by L\"uscher \cite{Lues} and the influence on the vacuum energy
by imposing periodic boundary conditions \(\phi(x+R)=\phi(x)\) is sometimes called topological Casimir effect \cite{Eli3, Toms}.
Originally the Casimir effect \cite{Casi} describes the phenomenon that a pair of perfectly conducting plates in the vacuum will feel a force
attracting each other. This can be understood as a response of the quantum fields on the influence of imposed boundary conditions, 
which in the case of 
parallel plates are Dirichlet boundary conditions (For recent review, see e.g. \cite{Bord}). In the case of periodic boundary conditions \cite{Ambj, Eli3}, e.g. studying quantum fields on a circle or a cylinder, there is not literally a material boundary. In this
sense we will use the terms 'ground state energy' and 'Casimir energy' synonymously.

The second step involves the {\it discretization} of the periodic space. One can expect that this will induce additional {\it finite-sites} effects
on the physical observables. Actor et. al.\cite{Acto} have investigated the Casimir effect on a lattice
in the case of Dirichlet boundary condition. Their treatment of the lattice was as an alternative regularization procedure compared to
mode number or momentum cutoff methods. Since their primary interest was in the continuum results, they ignored all 'lattice
quantities' because they 'take care of themselves' by vanishing for \(L\to\infty\). A numerical study on 
the influence of lattice Casimir effects on the dynamics of kinks in a discretized sine-Gordon model was done by Speight \cite{Spei}.

In the case of \((1+1)\)-dim relativistic integrable theories Zamolodchikov \cite{Zamo} suggested to use the thermodynamic Bethe ansatz (TBA)
\cite{Yang} by interpreting the ground state energy of the theory on a circle of circumference \(R\) as the free energy of the same theory on a
infinite line at temperature \(1/R\) \cite{Klas}. To apply this strategy on determining finite size-effects on the energy of strings in 
\(AdS_5\times S^5\) as advocated by Ambjorn et. al \cite{Amb2}, Arutyunov and Frolov \cite{Arut} have to introduce a mirror theory, since the 
gauge fixed worldsheet-theory has lost the explicit relativistic structure:
\begin{equation}\label{eq:mirr}
 H_{w.s.}=\sqrt{1+4g^2\sin^2\frac{p}{2}}\qquad\Longrightarrow \tilde H_{mirr}=2\,\mathrm{arsinh}\left(\frac{1}{2g}\sqrt{1+\tilde p^2}\right).
\end{equation}
For a review on the applications of TBA in the AdS/CFT context, see e.g. \cite{Jan2, Baj2}.

In this work we want to take the situation with a finite lattice structure and periodic boundary conditions for a scalar field at 
face value and ask, 
what kind of energy \(E_{Cas}(L,R,m)\) can be attributed to the corresponding 
lattice vacuum state. Any regularization or renormalization procedures, if necessary on a lattice at all, should take place for fixed \(L\). 
As we will argue in the course of this paper, we will indeed need a renormalization condition also on the lattice although all quantities
by construction are mathematical finite at first sight.

As ingredients we only use canonical quantization and zeta function regularization \cite{Kirs, Eliz}, adapted to the situation of a periodic lattice, to derive a compact analytic result for the Casimir energy. The expression can be interpreted as the free energy of particle at 
temperature \(1/L\) governed
by the mirror-typ Hamiltonian \(\tilde H_{mirr}\) in (\ref{eq:mirr}), where the 'coupling' constant is given by the dimensionless combination of the physical parameters \(g=L/(Rm)\).  From this we can extract analytic results for the finite-sites corrections
to the 'classic' Casimir energy \(E\sim 1/R\) in a large \(L\) expansion. Further, our result contains the
leading L\"uscher-typ corrections \(E\sim e^{-mR}\) and a 'wrapping'-type \cite{Amb2} correction \(E\sim g^{2L}\) as special cases.

We tried to make the paper self-contained as an lattice approach to the Casimir energy, but think it is necessary to point out some formal similarities with results discussed in the AdS/CFT context.
The plan of the paper is as follows. In Section 2 we review the general properties of lattice scalar fields and outline the shortcomings
of a naive identification of the zero-point energy, while in Section 3 we will use a careful treatment by introducing
the lattice zeta function associated to the lattice spectral problem to extract a physical sensible result for the lattice
Casimir energy. In Section 4 we demonstrate that our result contains the well established continuum Casimir energies for scalar fields as 
limiting cases and  a systematic large \(L\) expansion can be obtained. Finally in Section 5 we clarify some mathematical points which 
lie behind our physical result.

\section{Scalar field on the lattice}
For completeness we will review shortly the description of scalar fields on a spatial lattice in the Hamiltonian formulation \cite{Acto} 
and then discuss the corresponding ground state energy.

Consider a scalar field with mass \(m\) on a circle with circumference \(R=2\pi\), which is
governed by the Klein-Gordon equation
\begin{equation}\label{eq:KleinGordon}
 (\partial_{t}^2-\partial_{x}^2)\phi(t,x)+m^2\phi(t,x)=0, \qquad\qquad \phi(t,x+R)=\phi(t,x).
\end{equation}
A straightforward discretization of this second order equation on a periodic \(L\)-sites lattice is given by
\begin{equation}\label{eq:stringbitequation1}
 \ddot{\phi}_l-\frac{1}{\eta^2}\left[\phi_{l+1}-2\phi_l+\phi_{l-1}\right]+m^2\phi_l=0,\qquad\qquad l=0,...,L-1,\qquad \phi_{L}(t)=\phi_0(t).
\end{equation}
By taking the lattice spacing \(\eta=R/L\) to zero \(\eta\to0\) one gets back the continuous equation (\ref{eq:KleinGordon}).

The vacuum field configuration is just given by \(\phi^{(vac)}_l=0\) for all \(l=0,...,L-1\) and the fluctuations \(\Phi_l(t)\)
around this background field \(\phi_l(t)=\phi^{(vac)}_l +\Phi_l(t)\) will be treated as quantum fields in the canonical way.

Setting \(\Phi_l(t)=e^{i\Omega t}\Phi_l\) gives an eigenvalue problem for the difference equation
\begin{equation}\label{eq:differenceeq}
 -\frac{1}{\eta^2}\left[\Phi_{l+1}-2\Phi_l+\Phi_{l-1}\right]+m^2\Phi_l=\Omega^2\Phi_l,\qquad\qquad l=0,...,L-1,\qquad \Phi_{L}=\Phi_0.
\end{equation}
This problem is equivalent to diagonalizing the \(L\times L\)-matrix
%\begin{equation}\label{eq:diffmatrix}
% D=\frac{1}{\eta^2}\left(\begin{array}{ccccc} -2+m^2\eta^2 & 1 & 0 & ... & 1 \\ {} \\
% 1 & -2+m^2\eta^2 & 1 & ... & \\ {} \\
% ... & & & ... &\\ {} \\
% 0 & ... & 1 & -2+m^2\eta^2 & 1 \\ {} \\
% 1 & ... & 0 & 1 & -2+m^2\eta^2 \end{array}\right)
%\end{equation}
\begin{equation}\label{eq:diffmatrix}
 D=\frac{1}{\eta^2}\left(\begin{array}{ccccc} -2+m^2\eta^2 & 1 & 0 & ... & 1 \\
 1 & -2+m^2\eta^2 & 1 & ... & \\ 
 ... & & & ... &\\
 0 & ... & 1 & -2+m^2\eta^2 & 1 \\ 
 1 & ... & 0 & 1 & -2+m^2\eta^2 \end{array}\right)
\end{equation}
and can be solved by elementary methods of difference equation calculus \cite{Mick}, giving the fluctuation frequencies for this lattice fields as 
\begin{equation}\label{eq:massfreq}
 \Omega_n=\sqrt{\left(\frac{L}{\pi}\right)^2\sin^2\left(\frac{\pi n}{L}\right)+m^2}, \qquad n=0,...,L-1.
\end{equation}
The corresponding eigenmodes 
\begin{equation}
 \phi_{n,l}=\frac{1}{\sqrt{L}}e^{\frac{2\pi i}{L}nl}
\end{equation}
are normalized according to
\begin{equation}
 \sum_{l=0}^{L-1}\phi_{n,l}\phi_{n',l}^{*}=\frac{1}{\eta}\delta_{n,n'},\qquad\qquad 
 \sum_{n=0}^{L-1}\phi_{n,l}\phi_{n,l'}^{*}=\frac{1}{\eta}\delta_{l,l'}.
\end{equation}
The lattice analogs of the quantum field operator and its canonical conjugate momentum operator with \(\left[\Phi_l,\Pi_{l'}\right]=\frac{i}{\eta}\delta_{l,l'}\) can now be expanded in terms of the eigenmodes as
\begin{eqnarray}
 \Phi_l(t)=\sum_{n=0}^{L-1}\frac{1}{\sqrt{2\Omega_n}}\left[e^{-i\Omega_nt}\phi_{n,l}\hat a_n+
e^{i\Omega_nt}\phi_{n,l}^{*}\hat a^{\dagger}_n\right],\nonumber\\
 \Pi_l=-i\sum_{n=0}^{L-1}\sqrt{\frac{\Omega_n}{2}}\left[e^{-i\Omega_nt}\phi_{n,l}\hat a_n-e^{i\Omega_nt}\phi_{n,l}^*
 \hat a_n^{\dagger}\right],
\end{eqnarray}
where the creation and annihilation operators \(\hat a_{n}^{\dagger}\) and \(\hat a_n\) satisfy canonical commutation relations
\( [\hat a_n,\hat a_{n'}^{\dagger}]=\delta_{n,n'}\) and \([\hat a_n,\hat a_{n'}]=
 [\hat a_n^{\dagger},\hat a_{n'}^{\dagger}]=0\).

The lattice Hamiltonian 
\begin{equation}
 H=\frac{\eta}{2}\sum_{l=0}^{L-1}\left[\Pi_l\Pi_l+\Phi_l(D\Phi)_l\right]
\end{equation}
can then be written as
\begin{equation}\label{eq:Hamilton}
 H=\frac{1}{2}\sum_{n=0}^{L-1}\Omega_n\left[\hat a_n^{\dagger}\hat a_n+\hat a_n\hat a_n^{\dagger}\right]=
 \sum_{n=0}^{L-1}\Omega_n\left[\hat a_n^{\dagger}\hat a_n+\frac{1}{2}\right].
\end{equation}
The \(L\)-sites lattice vacuum \(|0\rangle_L\) is defined by the condition that it is destroyed by all annihilation operators:
\begin{equation}
 \hat a_n|0\rangle_L=0,\qquad \qquad n=0,\,1,...,L-1.
\end{equation}
Acting with the Hamiltonian (\ref{eq:Hamilton}) on this state gives a bare value \(E_0\) for the vacuum energy
\begin{equation}
 H|0\rangle_L=E_0(L,m)|0\rangle_L,
\end{equation}
with
\begin{equation}\label{eq:naivesum}
 E_0(L,m)=\frac{1}{2}\sum_{n=0}^{L-1}\Omega_n,
\end{equation}
where \(\Omega_n\) as in (\ref{eq:massfreq}).
Since the number of modes are finite and given by the number of sites \(L\), there seems to be no ambiguity to consider the sum (\ref{eq:naivesum})as the vacuum or ground state energy of the lattice model: \(E_{Cas}(L,m)\stackrel{?}{=}E_0(L,m)\).
Mathematically, the finite number \(L\) of frequencies gives a finite sum with a 
finite result. There are two main arguments why this prescription cannot give the right physical answer.

First of all, one expect that a lattice Casimir energy \(E_{Cas}(L,m)\) approaches for \(L\to\infty\) in a
smooth way the finite Casimir energy for periodic boundary conditions in the continuum as reported e.g. in \cite{Bord, Ambj}. 
This cannot be satisfied by (\ref{eq:naivesum}), since for \(L\to\infty\) it becomes the notorious divergent series,
which has to be renormalized.\footnote{In \cite{Acto} the corresponding sum (\ref{eq:naivesum}) for Dirichlet boundary conditions
was treated by the authors as a {\it lattice regularization} of the divergent sum in the continuum and not as the physical 
lattice Casimir energy.} 

Further, Casimir energies should be considered as a special case of quantum corrections to the energy of classical background fields 
\(\phi_{cl}\)\cite{Bor2}.  The background field in the case of the Casimir effect is just the vacuum 
field configuration \(\phi_{cl}=\phi^{(vac)}=0\) with energy \(E_{cl}(\phi_{cl})=0\). Only the fluctuations \(\phi_{fl}\) of the field 
\(\phi=\phi_{cl}+\phi_{fl}\) should be treated as quantum fields \(\phi_{fl}\to\Phi\), regardless whether \(\phi\) is a continuum field \(\phi(x)\) or a lattice field \(\phi_l\).

Now consider what happens with the frequencies (\ref{eq:massfreq}) in the limit of large 'mass' \(m\to\infty\):
\begin{equation}\label{eq:infinimass}
 \Omega_n\to m+\frac{1}{m}\frac{L^2}{2\pi^2}\sin^2\left(\frac{\pi n}{L}\right)-...%\frac{1}{m^3}\frac{L^4}{8\pi^4}
 %\sin^4\left(\frac{\pi n}{L}\right)+...
\end{equation}
In the large mass limit even the frequency of the lowest lying fluctuation mode escapes to infinity.
Since an infinite amount of energy is necessary to excite vacuum fluctuations it is physically reasonable, that no quantum fluctuations on 
top of the classical background field appear at all. Therefore one expect that also any quantum corrections to the energy vanish and 
one should be left only with the classical energy \(E=E_{cl}=0\). But a resummation of every term in (\ref{eq:infinimass}) according to 
(\ref{eq:naivesum}) gives
\begin{equation}\label{eq:wrongex}
 E_0\to \frac{1}{2}Lm+\frac{1}{m}\frac{L^3}{8\pi^2}-... ,%\frac{1}{m^3}\frac{3L^5}{128\pi^4}+...
\end{equation}
which implies \(E_0\to\infty\) for \(m\to\infty\). 

We see here very explicit that the demand for renormalization follows not from the appearance of a
divergent infinite sum at the very beginning. Although in the discretized version of the massive scalar field only a 
finite number of modes contribute, an infinity shows up which needs to be renormalized
\begin{equation}
 E_{Cas}=E_{ren}\to 0,\qquad m\to\infty.
\end{equation}
In the case of free scalars this corresponds to a specific normal ordered Hamiltonian
\(:H:\,=H-E_{Bulk}\), where \(E_{Bulk}\) should contain all terms which have be subtracted.

%Using instead 
% \begin{equation}
% E_{ren}\to 0,\qquad m, L\to\infty
%\end{equation}
%would imply that we have to discard all terms in (\ref{eq:wrongex}), which implies further that for no values of \(L\) and \(m\) 
%a non-zero Casimir energy appears. That is obviously wrong in the continuum limit \(L\to\infty\).

At least in the massless case \(m=0\) one can resum the original expression (\ref{eq:naivesum}) explicitly, which results in
\begin{equation}\label{eq:closedeq}
 E_0=\frac{L}{2\pi}\cot\left(\frac{\pi}{2L}\right).
\end{equation}
 Expanding this expression for \(L\to\infty\) gives
\begin{equation}
 E_0=\frac{L^2}{\pi^2}-\frac{1}{12}-\frac{\pi^2}{720L^2}+\mathcal{O}(L^{-3}).
\end{equation}
The first term is divergent in the continuum limit, the second term is the famous Casimir energy contribution,
the third term is an example for the sub-leading contributions to the energy coming from the discretization.
In the continuum limit we need a proper renormalization prescription to handle the divergent term proportional to \(L^2\) and
to end up with
\begin{equation}\label{eq:zeromass}
 E_{Cas}=\frac{L}{2\pi}\cot\left(\frac{\pi}{2L}\right)-\frac{L^2}{\pi^2},
\end{equation}
which gives for \(L\to\infty\)
\begin{equation}\label{eq:OneLoop1}
 E_{Cas}=-\frac{1}{12}.
\end{equation}

\section{Spectral zeta function for scalar fields on the lattice}
The previous discussion has shown that we will need a more careful analysis in order to determine the lattice Casimir energy.
In the following we will write \(\lambda_n=\Omega_n^2\) with \(\Omega_n\) as in (\ref{eq:massfreq}). In order to guarantee that no zero modes
are in the spectrum, we assume further \(m>0\).
To proceed we introduce the spectral zeta function for the corresponding difference operator (\ref{eq:differenceeq}) as
\begin{equation}
 \zeta_D(s)=\mu^{1+2s}\sum_{n=1}^{L-1}\lambda_n^{-s}
\end{equation}
and the 'regularized' energy is given by \footnote{Obviously, the regularized energy is a function of zeta function argument \(s\),  
the lattice sites number \(L\), and the mass parameter \(m\):
\(E_{reg}(s,L,m)\). In the following we will only mention the arguments explicitly, which are the most relevant ones in the course of arguing.}
\begin{equation}\label{eq:RegSum}
 E_{reg}(s)=\frac{1}{2}\zeta_D(s),
\end{equation}
where parameter \(\mu\) is introduced to keep track of dimensions.
The expression (\ref{eq:RegSum}) {\it is} finite for \(\mathrm{Re}(s)>0\) in the limit \(m\to\infty\) and for 
\(\mathrm{Re}(s)>\frac{1}{2}\) even in the limit \(L\to\infty\).
We emphasize, that we don't need the regularization because the sum over lattice frequencies is a divergent series. This is not the
case for the discrete field on a finite lattice. We need the regularization to handle the infinity in the large mass limit \(m\to\infty\).
In the previous section we have evaluated the zero-point energy as sum over the explicitly known fluctuation frequencies.
Now we will use an approach, which incorporate the fluctuation spectrum in a more intrinsic way, defined
as the roots \(\Delta(\lambda)-2=0\) of the spectral discriminant
\begin{equation}\label{eq:specdis}
 \Delta(\lambda)=2\cosh\left[L\,\mathrm{arcosh}\left(1-\frac{2\pi^2}{L^2}(\lambda-m^2)\right)\right].
\end{equation}
Then we can rewrite the spectral zeta function in terms of a contour integral representation \cite{Kirs} as
\begin{equation}
 \zeta_D(s)=\frac{1}{2\pi i}\mu^{1+2s}\int_{\gamma}\mathrm{d}\lambda\,\lambda^{-s}\frac{\partial}{\partial\lambda}\ln(\Delta(\lambda)-2)=
 \frac{1}{2\pi i}\mu^{1+2s}\int_{\gamma}\mathrm{d}\lambda\,\lambda^{-s}R(\lambda),
\end{equation}

The resolvent 
\begin{equation}
 R(\lambda)=\frac{\partial}{\partial\lambda}\ln(\Delta(\lambda)-2)=\frac{\Delta'(\lambda)}{\Delta(\lambda)-2}
\end{equation}
has therefore \(L\) poles on the positive real axis at the places of the eigenfrequencies and the integration contour \(\gamma\) encircles them. 
The advantage of this representation is, that it can be applied also in cases where the fluctuation spectrum
is not known. The only requirement is that one has a spectrum defining equation, such as the spectral discriminant in (\ref{eq:specdis}).

Deforming the contour to lie along the branch cut on the negative real axis \(\lambda\in [-\infty,0]\), one arrives at the following result:
\begin{equation}
 E_{reg}(s)=E_{Bulk}(s)+E_{Cas}(s)
\end{equation}
with
\begin{eqnarray}\label{eq:splitint}
 E_{Bulk}(s)&=&L\frac{\sin(\pi s)}{2\pi}\mu^{1+2s}\int_0^{\infty}\mathrm{d}\lambda\,\lambda^{-s}\frac{\partial}{\partial\lambda}q(-\lambda),
 \nonumber\\
 E_{Cas}(s)&=&\frac{\sin(\pi s)}{\pi}\mu^{1+2s}\int_0^{\infty}\mathrm{d}\lambda\,\lambda^{-s}\frac{\partial}{\partial\lambda}\ln\left(1-e^{-Lq(-\lambda)}\right),
\end{eqnarray}
and
\begin{equation}
 q(\lambda)=\mathrm{arcosh}\left(1-\frac{2\pi^2}{L^2}(\lambda-m^2)\right).
\end{equation}

The splitting of the integral into two parts is valid for \(0<\mathrm{Re}(s)<1\). Since from the behaviour of the integrands one can estimate that
the integral in \(E_{Bulk}(s)\) is convergent for \(0<\mathrm{Re}(s)<1\) 
and the integral in \(E_{Cas}(s)\) is convergent for \(\mathrm{Re}(s)<1\).

Inserting 
\begin{equation}\label{eq:ZeroGapDiff}
 \frac{\partial}{\partial\lambda}q(-\lambda)=\frac{\pi}{L}\frac{1}{\sqrt{(\lambda+m^2)(1+\frac{\pi^2}{L^2}(\lambda+m^2))}}
\end{equation}
into the integrand of \(E_{Bulk}(s)\)
and introducing the new integration variable \(x^2=\lambda+m^2\) 
the integral can be solved for \(\mathrm{Re}(s)>0\) in terms of a hypergeometric function:
\begin{equation}\label{eq:BulkResult}
 E_{Bulk}(s)=\frac{1}{2}m^{-2s}\mu^{1+2s}L\,{}_2F_1\left(s,\,\frac{1}{2},\,1;\,-\frac{L^2}{\pi^2m^2}\right).
\end{equation}
Although the integral in (\ref{eq:splitint}) was divergent for \(s<0\), we can use the result (\ref{eq:BulkResult}) as analytic continuation 
of \(E_{Bulk}(s)\) to negative values of \(s\) and evaluate it also for \(s=-\frac{1}{2}\):
\begin{equation}\label{eq:LatticeDiv}
 E_{Bulk}(-1/2)=\frac{Lm}{\pi}\mathbb{E}\left(-\frac{L^2}{\pi^2m^2}\right),
\end{equation}
where \(\mathbb{E}\) is the complete elliptic integral of second kind.
Now we expand for \(L\to\infty\) and get
\begin{equation}
 E_{Bulk}(-1/2)=\frac{L^2}{\pi^2}+E_{Bulk}^{(0)}+\frac{1}{L^2}E_{Bulk}^{(2)}+\frac{1}{L^4}E_{Bulk}^{(4)}+...
\end{equation}
with
\begin{eqnarray}
  E_{Bulk}^{(0)}&=&\frac{m^2}{2}\left[\frac{1}{2}+2\ln2+\ln\frac{L}{\pi}-\ln m\right],\\
  E_{Bulk}^{(2)}&=&\frac{\pi^2m^4}{16}\left[\frac{3}{4}-2\ln2-\ln\frac{L}{\pi}+\ln m\right].\\
%  E_{Bulk}^{(4)}&=&\frac{3\pi^4m^6}{128}\left[-1+2\ln2+\ln\frac{L}{\pi}-\ln m\right]
\end{eqnarray}
In general, the terms \(E_{Bulk}^{(2n)}\) are \(\mathcal{O}(m^{2n+2})\).
From these expansions one can see, that \(E_{Bulk}\) encapsulates the terms, which diverge for \(m\to\infty\) and/or \(L\to\infty\).
Applying the large mass subtraction scheme the renormalized lattice Hamiltonian (\ref{eq:Hamilton}) should be defined by
\begin{equation}
 :H:\, = H -E_{Bulk}(-1/2,L,m).
\end{equation}
Acting on the vacuum state \(|0\rangle_L\) provides now
\begin{equation}
 :H:|0\rangle_L=E_{Cas}(L,m)|0\rangle_L,
\end{equation}
where \(E_{Cas}(L,m)\) is given in (\ref{eq:splitint}).

This renormalization ensures that for any given values \(L\) of lattice sites the ground state energy in
the limit \(m\to\infty\) vanishes and, as we will see in the next section, this choice is also compatible with the results known from the continuum
theory.

There are several integral representations for \(E_{Cas}(L,m)\) possible.
After integration by parts one can write the Casimir energy of a free scalar field
with mass \(m\) on a periodic lattice with \(L\) sites as
\begin{equation}\label{eq:FinalCasimir}
 E_{Cas}(L,m)=\frac{1}{\pi}\int_{m}^{\infty}\mathrm{d}\kappa\,\tilde p'(\kappa)\ln\left(1-e^{-\varepsilon(\kappa)}\right),
\end{equation}
with
\begin{equation}\label{eq:moment}
 \tilde p(\kappa)=\sqrt{\kappa^2-m^2},\qquad \varepsilon(\kappa)=2L\,\mathrm{arsinh}\left(\frac{\pi}{L}\kappa\right).
\end{equation}

A more illuminating representation can be obtained by a minor rescaling of the integration variable \(\tilde p=m\bar p\) and introducing the dimensionless parameter \(g=\frac{L}{Rm}\). Then one can write the Casimir energy more precisely as
\begin{equation}
 E_{Cas}(L,g,m)=2m\,f(L,g),
\end{equation}
where the dimensionless scaling function
\begin{equation}\label{eq:freeen}
 f(L,g)=\int_0^{\infty}\frac{\mathrm{d}\bar p}{2\pi}\ln\left(1-e^{-L\tilde H_g(\bar p)}\right)
\end{equation}
can be interpreted as thermodynamic free energy of particles on an infinite line at a temperature \(\frac{1}{L}\) governed by the Hamiltonian
\begin{equation}
 \tilde H_g(\bar p)=2\,\mathrm{arsinh}\left(\frac{1}{2g}\sqrt{\bar p^2+1}\right).
\end{equation}
This rather exotic looking Hamiltonian is formally exactly the same as the one in the so called
mirror theory, introduced by Arutyunov and Frolov by a double Wick rotation of the light-cone gauge-fixed \(AdS_5\times S^5\) worldsheet theory \cite{Arut}. The emergence of this Hamiltonian related to a lattice model is not as surprising as it may seem at first sight. The eigenfrequencies (\ref{eq:massfreq}) of the lattice field modes mimic the dispersion relation \(H_{w.s}\) in (\ref{eq:mirr}) obtained for the string worldsheet modes after light-cone gauge-fixing 
and are also similar to the dispersion relation of giant magnons, solitonic states propagating
on the worldsheet \cite{Hofm}. This similarity had previously led to musings about an underlying lattice structure of the \(AdS_5\times S^5\) 
worldsheet \cite{Zare}.  

\section{Discussion}

The spectrum of free scalars with bare mass \(m\) on a \(L\)-sites periodic lattice with circumference \(R\) starts
 with (\(g=L/(mR)\))
\begin{itemize}
 \item the vacuum state 
 \begin{equation}
  |0\rangle_L,\qquad\qquad E=E_{Cas}=2m\,f(L,g),
 \end{equation}
 \item the one-particle states 
 \begin{equation}
  \hat a_n^{\dagger}|0\rangle_L,\qquad \qquad E=\sqrt{\left(\frac{L}{\pi}\right)^2\sin^2\left(\frac{2\pi n}{L}\right)+m^2}+2m\,f(L,g), \qquad
 n=0,...,L-1
 \end{equation}
\end{itemize}

By investigating certain limits in the dimensionless parameters \(L\) and \(g\), we are able to show
that our result (\ref{eq:freeen}) contains some well known as well as some new results as special cases.

\begin{figure}
 \includegraphics[scale=0.8]{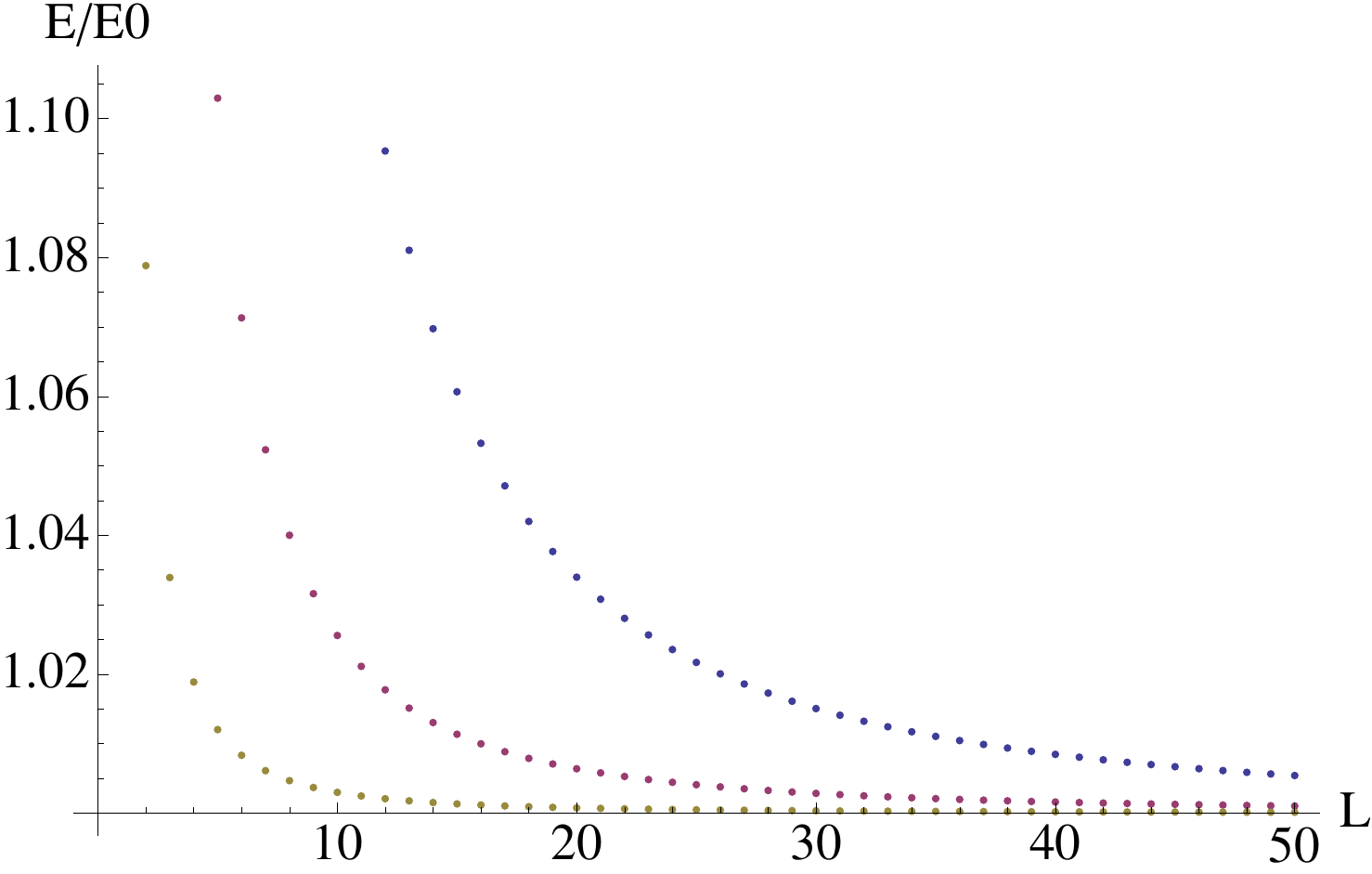}
 \caption{Relative deviation of the lattice Casimir energy \(E_{Cas}(L,2\pi,m)\) compared to the continuum result \(E^{(0)}(2\pi,m)\), evaluated for 
 \(m=9/10,\, 1/2\) and \(1/10\) (blue, magenta and yellow dots, respectively) as a function of lattice sites \(L\). 
 This shows that lattice effects are under control also for a small lattice as long as \(mR<1\).}
\end{figure} 

%\begin{figure}
% \includegraphics[scale=0.7]{CasimirPlot2.pdf}
% \caption{Casimir energy for \(m=\frac{1}{2}\)as function of lattice sites \(L\)}
%\end{figure} 

\begin{figure}
 \includegraphics[scale=0.7]{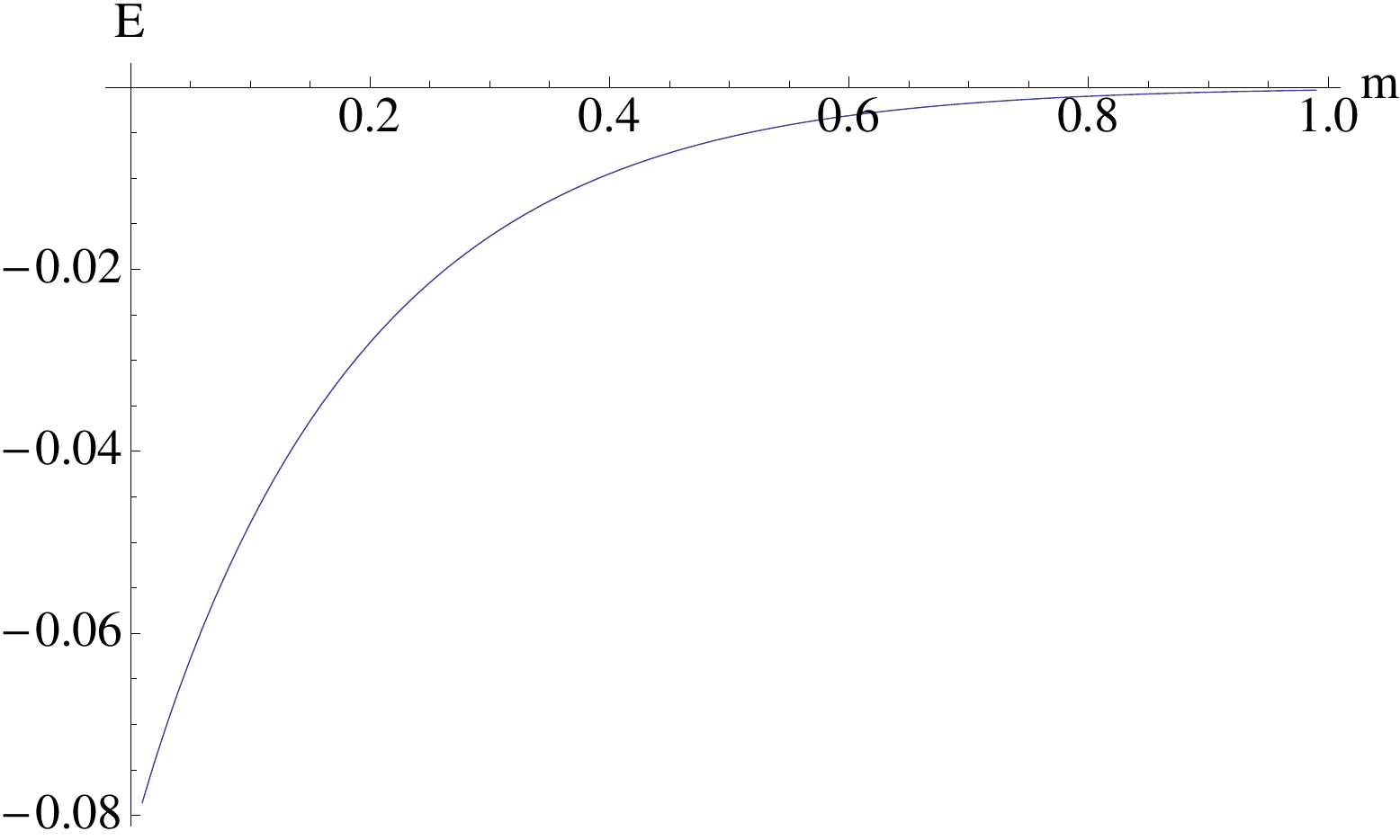}
 \caption{Casimir energy depending on mass parameter \(m\) for fixed lattice \(L=30\)}
\end{figure}

\subsection{The continuums limit}

The limit  \(L\), \(g\to\infty\), while \(Rm=L/g\) held fixed, corresponds to the continuum limit of a scalar field on a compact circle with
circumference \(R\). From (\ref{eq:freeen}) one obtains an inverse power-law behaviour in large \(L\):
\begin{equation}
 E_{Cas}(L,R,m)\to E^{(0)}(R,m)+\frac{1}{L^2}E^{(2)}(R,m)+\frac{1}{L^4}E^{(4)}(R,m)+\mathcal{O}(L^{-6}),
\end{equation}
where the first coefficients have the following integral representations:
\begin{eqnarray}\label{eq:largelimit}
 E^{(0)}(R,m)&=&\frac{m}{\pi}\int_0^{\infty}\mathrm{d}\bar p\,\ln\left(1-e^{-Rm\sqrt{\bar p^2+1}}\right), \nonumber\\
 E^{(2)}(R,m)&=&-\frac{R^3}{24\pi}\int_{m}^{\infty}\mathrm{d}\kappa\frac{\kappa^4}{\sqrt{\kappa^2-m^2}(e^{R\kappa}-1)},\nonumber\\
 E^{(4)}(R,m)&=&\frac{R^5}{5760\pi}\int_{m}^{\infty}\mathrm{d}\kappa\frac{\kappa^6}{\sqrt{\kappa^2-m^2}}\left[\frac{27}{e^{R\kappa}-1}-
 \frac{2R\kappa e^{R\kappa}}{(e^{R\kappa}-1)^2}\right].
\end{eqnarray}

\begin{figure}
 \includegraphics[scale=0.8]{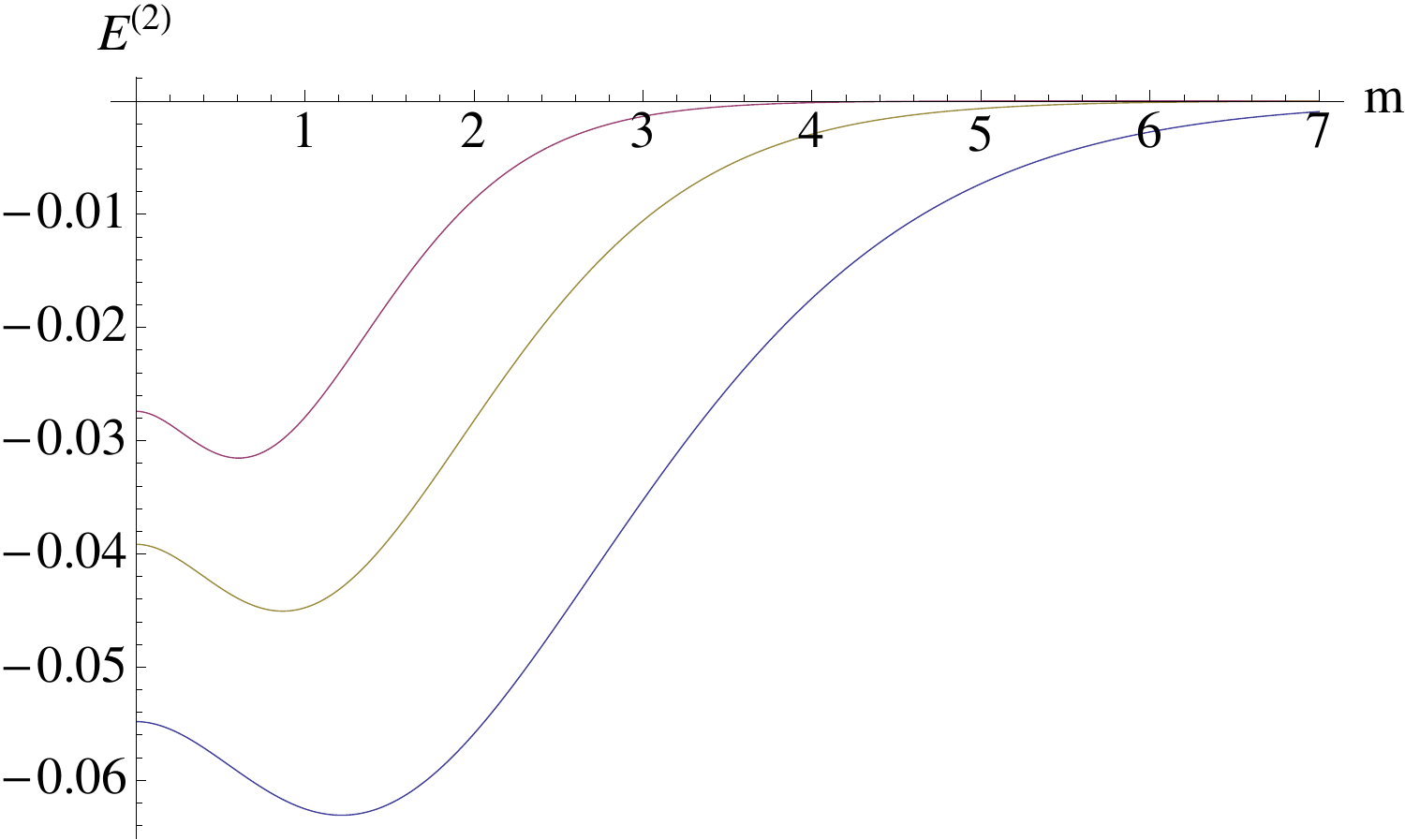}
 \caption{The sub-leading lattice correction \(E^{(2)}(R,m)\) for fixed circumference \(R=\frac{\pi}{2},\frac{7}{10}\pi\) 
and \(\pi\) in magenta, blue and yellow.}
\end{figure} 

\begin{figure}
 \includegraphics[scale=0.8]{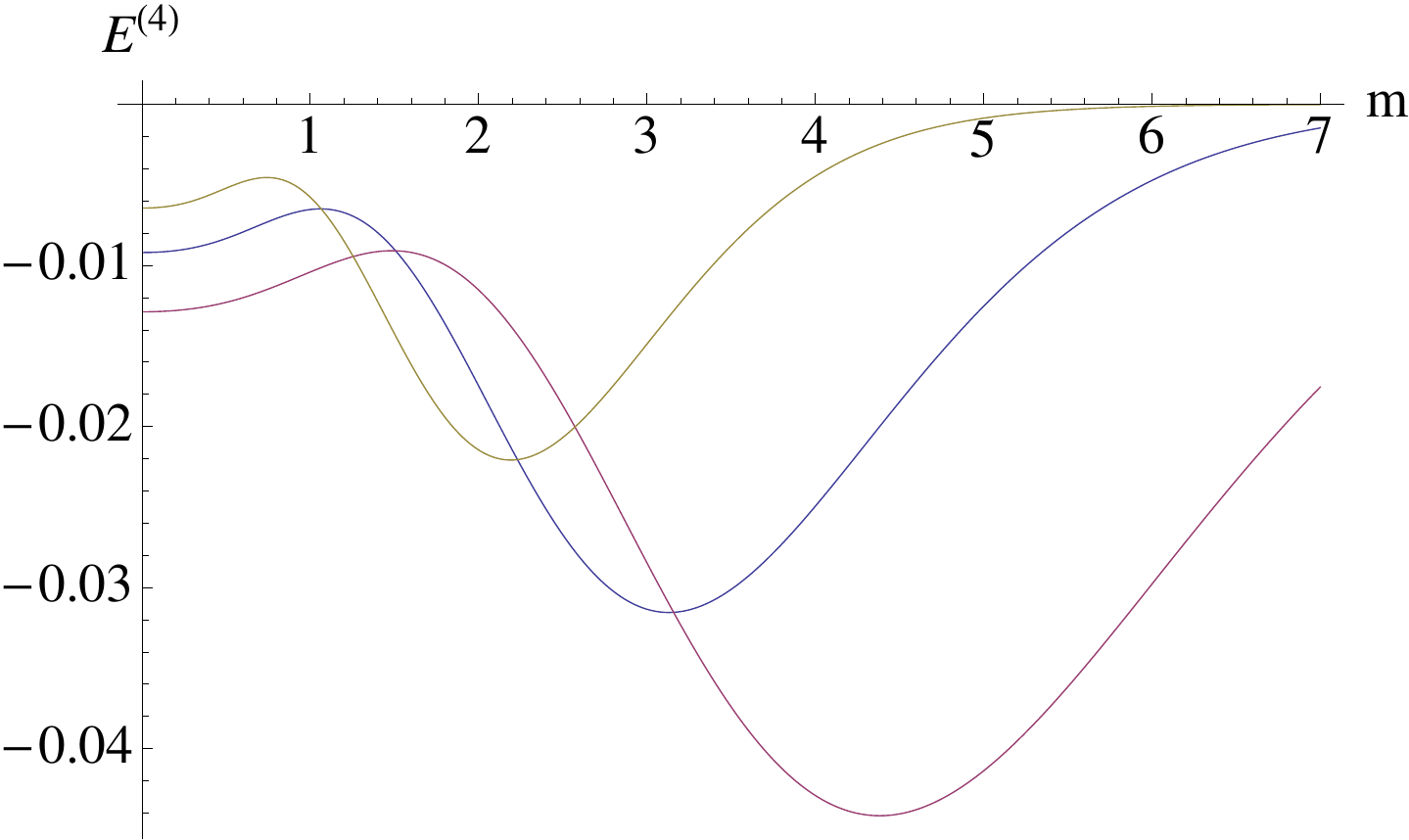}
 \caption{The sub-sub-leading lattice correction \(E^{(4)}(R,m)\) for fixed circumference \(R=\frac{\pi}{2},\frac{7}{10}\pi\) 
 and \(\pi\) in magenta, blue and yellow.}
\end{figure} 

For the leading order contribution \(E^{(0)}\) it is preferable to introduce in the integral representation a rapidity parameterization
\footnote{There exist also the possibility to introduce an elliptic parameterization of the integrand in (\ref{eq:freeen}) in
terms of Jacobi elliptic functions. But it seems, this will not lead to an integration in terms of 'classic higher transcendental functions' such as
the modified Bessel function \(K_{\nu}(z)\) in the case of the hyperbolic parameterization.}
\(\bar p=\sinh(\theta)\). Then one can see that this term can be interpreted as the free energy of an ideal relativistic
Bose gas \cite{Klas} at temperature \(1/R\)
\begin{equation}
 E^{(0)}(R,m)=\frac{m}{\pi}\int_0^{\infty}\mathrm{d}\theta\cosh(\theta)\ln\left(1-e^{-mR\cosh(\theta)}\right),
\end{equation}
as well as the Casimir energy of a massive scalar field on a circle of circumference \(R\) with periodic boundary conditions \cite{Bord, Ambj}
\begin{equation}\label{eq:E0}
 E^{(0)}(R,m)=-\frac{m}{\pi}\sum_{n=1}^{\infty}\frac{1}{n}K_1(mR n).
\end{equation}
Further, in the limit \(R\to\infty\) it reproduce the leading L\"uscher corrections \cite{Lues, Bajn}:
\begin{equation}
 E^{(0)}(R,m)\stackrel{R\to\infty}{\longrightarrow}
-\frac{m}{\pi}\int_0^{\infty}\mathrm{d}\theta\cosh(\theta)e^{-mR\cosh(\theta)}=-\frac{m}{\pi}K_1(mR).
\end{equation}
The sub-leading term in (\ref{eq:largelimit}) provides the first lattice correction of \(\mathcal{O}(L^{-2})\) and can also be expressed
in terms of modified Bessel functions:
\begin{equation}\label{eq:E2}
 E^{(2)}(R,m)=-\frac{m^2R}{24\pi}\left[3\sum_{n=1}^{\infty}\frac{1}{n^2}K_2(mR n)-6mR
 \sum_{n=1}^{\infty}\frac{1}{n}K_3(mR n)+m^2R^2\sum_{n=1}^{\infty}K_4(mR n)\right].
\end{equation}

The general term \(E^{(n)}(2\pi,m)\) in the large \(L\) expansion is composed of integrals of the type 
\begin{equation}\label{eq:BesselInts}
 I(a,b,c,d,R,m)=\int_m^{\infty}\mathrm{d}\kappa(\kappa^2-m^2)^{\frac{a}{2}}\kappa^b\frac{(e^{R\kappa})^c}{(e^{R\kappa}-1)^d}.
\end{equation}
These integrals can be expressed in terms of modified Bessel functions and their properties are summarized in Appendix A.

The sub-sub-leading lattice contributions \(E^{(4)}\) is given in terms of (\ref{eq:BesselInts}) as
\begin{equation}
 E^{(4)}(R,m)=\frac{R^5}{5760\pi}\left[27I(-1,6,0,1,R,m)-2R\,I(-1,7,1,2,R,m)\right].
\end{equation}
By iterative application of the reduction formula (\ref{eq:reduceformula}) the final expression in terms of Bessel function series
will be of the following type:
\begin{equation}
 E^{(4)}(R,m)=\sum_{\nu=3}^7\sum_{n=1}^{\infty}c_{\nu}(R,m)\alpha_{\nu}(n)K_{\nu}(mR n),
\end{equation}
with \(\alpha_{\nu}(n)\) a rational function in \(n\) and prefactors \(c_{nu}(R,m)\).

In Fig.1 we have examined the relative deviation of the lattice Casimir energy \(E_{Cas}(L,2\pi,m)\) to the corresponding continuum result
 \(E^{(0)}(2\pi,m)\), which shows that for \(mR<1\) the lattice corrections are under control also for small
lattices \(L<10\). 

In Fig.3 and Fig.4 the sub-leading lattice contributions \(E^{(2)}\) and \(E^{(4)}\) are plotted as function of the mass \(m\) for
three different values of the compactification size \(R=\pi/2, 7\pi/10\) and \(\pi\). Although the values are roughly of the same order 
of magnitude, one should keep in mind that these contribution are suppressed by \(1/L^2\) and \(1/L^4\), respectively.

\subsection{The massless limit \(m\to 0\)}
The limit \(g\to\infty\) and \(L\) held fixed corresponds to the limit of massless scalars on the lattice.
We can immediately obtain the corresponding Casimir energy by setting \(m=0\) in (\ref{eq:FinalCasimir}):
\begin{equation}\label{eq:Caszeromass}
 E_{Cas}(L)=\frac{1}{\pi}\int_0^{\infty}\mathrm{d}\kappa\ln\left(1-e^{-2L\mathrm{arsinh}\left(\frac{\pi}{L}\kappa\right)}\right).
\end{equation}
By construction, this should be equal to the previous mentioned result (\ref{eq:zeromass})
\begin{equation}\label{eq:zeromass2}
 E_{Cas}(L)=\frac{L}{2\pi}\cot\left(\frac{\pi}{2L}\right)-\frac{L^2}{\pi^2}
\end{equation}
with continuums limit \(L\to\infty\) expansion as
\begin{equation}\label{eq:masslessexp}
 E_{Cas}\to -\frac{1}{12}-\frac{\pi^2}{720}\frac{1}{L^2}+\mathcal{O}(L^4).
\end{equation}

On the other hand we can evaluate the massless limit from the expressions (\ref{eq:E0}) and (\ref{eq:E2}), which were obtained by taking 
first the limit \(L\to\infty\):
\begin{eqnarray}
 E_{ren}^{(0)}&\to&-\frac{1}{\pi R}\sum_{n=1}^{\infty}\frac{1}{n^2}=-\frac{1}{\pi R}\zeta(2)=-\frac{\pi}{6R},\\
 E_{ren}^{(2)}&\to& -\frac{m^2R}{24\pi}\left[\frac{6}{R^2m^2}-\frac{48}{m^2R^2}
 +\frac{48}{m^2R^2}\right]\sum_{n=1}^{\infty}\frac{1}{n^4}=-\frac{1}{4\pi R}\zeta(4)=-\frac{\pi^3}{360 R},
\end{eqnarray}
which is in perfect agreement with (\ref{eq:masslessexp}) for \(R=2\pi\). Notice also the appearance of Riemann zeta function \(\zeta(s)\).

Although (\ref{eq:FinalCasimir}) was derived under the assumption \(m>0\) in order to avoid zero modes in the fluctuation spectrum, 
we can treat \(m\) now just as an infrared regulator which we removed \(m\to 0\) in the final step towards (\ref{eq:Caszeromass}).
Further we recognized that for an numerical evaluation of (\ref{eq:Caszeromass}) it is advantageous to use the logarithmic representation
of the inverse hyperbolic function:
\begin{equation}
 E_{Cas}(L)=\frac{1}{\pi}\int_0^{\infty}\mathrm{d}\kappa\ln\left(1-\left(\pi\kappa/L+
 \sqrt{\left(\pi\kappa/L\right)^2+1}\right)^{-2L}\right).
\end{equation}

\subsection{Decompactified lattice limit}
There are two different limit procedures, which result in a decompactification \(R\to\infty\).
\begin{itemize}
\item
Consider the limit \(L\to\infty\) while \(g\) held fixed. This corresponds to \(R, L\to\infty\) while \(\eta=R/L\) held fixed, which is the
situation of an infinite lattice with constant lattice spacing \(\eta\). The Casimir energy is
\begin{equation}
 E_{Cas}(L,g)\to-2m\sum_{n=1}^{\infty}\frac{1}{n}\int_0^{\infty}\frac{\mathrm{d}\bar p}{2\pi}
e^{-2nL\,\mathrm{arsinh}\left(\frac{1}{2g}\sqrt{\bar p^2+1}\right)}.
\end{equation}
Clearly, in this situation the lattice corrections are exponentially suppressed \(\mathcal{O}(e^{-L})\) and vanish completely in the
strict \(L\to\infty\) limit.
\item 
The other decompactification limit can be obtained by taking \(g\to 0\) and \(L\) held fixed, which corresponds to a 'dilute lattice' with 
infinite large lattice spacing \(\eta\to\infty\).
The Casimir function can be approximated in this case as
\begin{equation}
  E_{Cas}(L,g)\to -2mg^{2L}\int_0^{\infty}\frac{\mathrm{d}\bar p}{2\pi}\frac{1}{(\bar p^2+1)^L}e^{-\frac{2Lg^2}{\bar p^2+1}+\mathcal{O}(g^4)}.
\end{equation}
Therefore the first contributions of the Casimir energy on an dilute \(L\)-sites lattice are delayed up to order \(g^{2L}\).
Interestingly, this coincide with the behaviour of wrapping contributions to anomalous dimensions of certain \(N=4\) SYM gauge operators, which
are expected not to be captured by standard spin chain Bethe ansaetze \cite{Amb2, Bajn}. 

\end{itemize}

\subsection{The roadmap}
The different limit procedures can be summarized in the following roadmap, where we also sketch the qualitative behaviour of the
vacuum energy in the major branches. Starting point is a massive scalar field on a periodic lattice of any size \(R\) with an arbitrary number of sites \(L\):

\begin{diagram}
  &   &   &                 &  \begin{array}{c} \text{finite periodic lattice} \\ \text{eq. (45), (46)}\end{array} &   &  &   & \\
  &   &   & \ldImplies^{L,g\to\infty}_{\eta\to 0} &  & \rdImplies^{R\to\infty} &  &   &  \\
         &           &  \begin{array}{c} \text{continuum on circle} \\  E\sim E^{(0)}(R)+L^{-2}E^{(2)}(R)+...\end{array}&            &         &            & 
\begin{array}{c} \text{lattice on infinite line} \\  \end{array} &            & \\
         & \ldImplies^{m\to 0} &              & \rdImplies_{R\to\infty} &         & \ldImplies_{g\to\infty}^{\eta\to 0} &          & 
\rdImplies^{g\to 0}_{\eta\to\infty}  &  \\
\begin{array}{c}\text{Casimir}\\ E\sim \frac{1}{R}\end{array} &            &              &            & \begin{array}{c}
\text{L\"uscher} \\ E\sim e^{-mR}\end{array}             &            &            &   & \begin{array}{c} \text{'Wrapping'} \\ 
E\sim g^{2L}\end{array}
\end{diagram} 

\section{Comments}
The observation that the lattice Casimir energy has a non-zero value at all can be traced back mathematically to the following issue.
For \(s=-1, -2, -3,...\) the lattice spectral zeta function
\begin{equation}\label{eq:doublesum}
 \zeta_D(s,L,m)=\sum_{n=0}^{L-1}\left[\left(\frac{L}{\pi}\right)^2\sin^2\left(\frac{\pi n}{L}\right)+m^2\right]^{-s}=
 m^{-2s}\sum_{n=0}^{L-1}\sum_{k=0}^{-s}\binom{-s}{k}\left(\frac{L}{\pi m}\right)^{2k}
 \sin^{2k}\left(\frac{\pi n}{L}\right)
\end{equation}
has a finite binomial expansion and there is no harm to interchange the two finite summations.
Using 
\begin{equation}
 \sum_{n=0}^{L-1}\sin^{2k}\left(\frac{\pi n}{L}\right)=\binom{2k}{k}\frac{L}{2^{2k}},\qquad k\in\mathbb{N},
\end{equation}
gives then the following alternative representation of the lattice spectral zeta function in terms of a Gaussian hypergeometric function
\begin{equation}\label{eq:zetaresum}
 \zeta_D(s,L,m)=Lm^{-2s}\,{}_2F_1\left(s,\frac{1}{2},1;-\frac{L^2}{\pi^2m^2}\right),\qquad s=-1,\,-2,\,-3,...
\end{equation}
which is essentially a polynomial of order \(-s\) in the argument \(-\frac{L^2}{\pi^2m^2}\). That the result should be a polynomial can 
also be seen from the fact that for \(s\) being a negative integer the zeta function is just the trace of the corresponding matrix \(D\) given in 
(\ref{eq:diffmatrix}) and taken to the power \(-s\): 
\begin{equation}
 \zeta_D(s,L,m)=\mathrm{Tr}\left(D^{-s}\right).
\end{equation}

Nevertheless, as the discussion of the previous section has shown, for general \(\mathrm{Re}(s)<1\) and \(s\neq -1, -2,...\) the identity (\ref{eq:zetaresum}) does not hold and should be corrected as
\begin{equation}
 \zeta_D(s,L,m)=Lm^{-2s}\,{}_2F_1\left(s,\frac{1}{2},1;-\frac{L^2}{\pi^2m^2}\right)+ \delta_D(s,L),
\end{equation}
where the additional correction term for \(\mathrm{Re}(s)<1\) has been found as
\begin{equation}\label{eq:errterm}
 \delta_D(s,L,m)= \frac{\sin(\pi s)}{\pi}\int_m^{\infty}\mathrm{d}x(x^2-m^2)^{-s}\frac{\partial}{\partial x}
 \ln\left(1-e^{-L\mathrm{arcosh}\left(1+\frac{2\pi^2}{L^2}x^2\right)}\right).
\end{equation}
For general values of \(s\) it is not allowed to interchange the double sum in (\ref{eq:doublesum}).\footnote{For a similar
issue in the case of double series representations of Epstein zeta functions, see \cite{Eli2}.} If one performs this
changing regardless better knowledge one has to correct the induced error by the additional term (\ref{eq:errterm}), which
for \(s=-\frac{1}{2}\) is essentially the Casimir contribution \(\delta_D(-1/2,L,m)=2E_{Cas}(L,m)\).

\section{Summary}
The aim of this paper was to derive an unambiguous result for the ground state energy {\it aka} Casimir energy of scalar fields on a \(L\)-sites 
periodic lattice. From the physics perspective we have pointed out that although all relevant quantities are finite on a lattice we need a renormalization prescription to extract a physical meaningful result for the lattice Casimir energy. We have argued that
in order to have consistency with established results for the ground state energy in the continuum, one has
to interpret the Casimir energy also on the finite lattice as quantum correction to the energy of a background field configuration. 
In the case of free fields this results in an \(L\)-dependent normal ordering prescription of the lattice Hamiltonian.

The final result for the Casimir energy can also be read as the thermodynamic free energy of particles at temperature \(1/L\) on a continuous line governed by the Arutyunov-Frolov mirror Hamiltonian \cite{Arut}. This interpretation indicates that the considerations on
ground state energies in (1+1)-dim relativistic field theories on a circle outlined by Zamolodchikov \cite{Zamo} seems to be valid
at least for free fields in a strict lattice description. Further, it is a strong support for our choice of finite renormalization of the lattice
Hamiltonian, that in several limiting cases well established results such as L\"uscher corrections, massive Casimir energy or
the so called 'wrapping' contributions are obtained automatically.
 
As mathematical origin of a non-zero lattice Casimir energy we have identified the issue of non-interchangeability of a double sum in the defining representation of the lattice spectral zeta function. The contour integral representation delivers an elegant way to determine this
remainder term, cf. this result with an attempt to use an Euler-Maclaurin-type formula. A further advantage of the contour integral representation of the spectral zeta function used in this paper is, that it can in principle also be applied in situations where the fluctuation spectrum is not explicit known. It would be interesting to apply this strategy to lattice models with
interacting fields, which allow for kink states \cite{Spei}.

\begin{acknowledgments}
 This work was supported by the Swedish Research Council (VR) under  contract no. 621-2010-3708.
\end{acknowledgments}

\appendix
\section{Bessel functions identities}

We consider integrals of the type 
\begin{equation}\label{eq:Besselints}
 I(a,b,c,d,R,m)=\int_m^{\infty}\mathrm{d}\kappa(\kappa^2-m^2)^{\frac{a}{2}}\kappa^b\frac{(e^{R\kappa})^c}{(e^{R\kappa}-1)^d}
\end{equation}
with parameters \(a\), \(b\), \(c\) and \(d\) chosen in order to have convergence.

Some basic examples of these integrals includes 
\begin{eqnarray}
 I(-1,0,0,1,R,m)&=&\sum_{n=1}^{\infty}K_0(mR n),\\
 I(1,0,0,1,R,m)&=&\frac{m}{R}\sum_{n=1}^{\infty}\frac{1}{n}K_1(mR n).%\\
 %I(a,0,0,1,m)&=&\frac{m^{\frac{a+1}{2}}}{\pi^{\frac{a+2}{2}}}\Gamma\left(\frac{a+2}{2}\right)\sum_{n=1}^{\infty}\frac{1}{n^{\frac{a+1}{2}}}
 %K_{\frac{a+1}{2}}(2\pi n m),\\
 %I(a,1,0,1,m)&=&\frac{m^{\frac{a+3}{2}}}{\pi^{\frac{a+2}{2}}}\Gamma\left(\frac{a+2}{2}\right)\sum_{n=1}^{\infty}\frac{1}{n^{\frac{a+1}{2}}}
 %K_{\frac{a+3}{2}}(2\pi n m)
\end{eqnarray}

In general, any convergent integral of type (\ref{eq:Besselints}) with \(b=0\) or \(b=1\) is expressible as a series of Bessel functions:
\begin{eqnarray}\label{eq:basicints}
 I(a,0,c,d,R,m)&=&\frac{m^{a+2}2^{\frac{1+a}{2}}}{\sqrt{\pi}(mR)^{\frac{1+a}{2}}}\Gamma\left(\frac{a+2}{2}\right)
 \sum_{n=d-c}^{\infty}(-1)^n\binom{-d}{n+d-c}\frac{1}{n^{\frac{a+1}{2}}}K_{\frac{a+1}{2}}(mR n),\nonumber\\ {} \\
 I(a,1,c,d,R,m)&=&\frac{m^{a+2}2^{\frac{1+a}{2}}}{\sqrt{\pi}(mR)^{\frac{1+a}{2}}}\Gamma\left(\frac{a+2}{2}\right)
 \sum_{n=d-c}^{\infty}(-1)^n\binom{-d}{n+d-c}\frac{1}{n^{\frac{a+1}{2}}}K_{\frac{a+3}{2}}(mR n). \nonumber
\end{eqnarray}
By partial integration one can show the following algebraic relation between different integrals (\ref{eq:Besselints})
\begin{eqnarray}\label{eq:reduceformula}
 (a+2)I(a,b,c,d,m)&=&-(b-1)I(a+2,b-2,c,d,m)-Rc\,I(a+2,b-1,c,d,m)+\nonumber\\
 & &+Rd\,I(a+2,b-1,c+1,d+1,m).
\end{eqnarray}
Therefore any (\ref{eq:Besselints}) with \(b>1\) can be reduced by a finite number of iterative applications of (\ref{eq:reduceformula})
to a linear combination of Bessel function series given in (\ref{eq:basicints}).

\end{document}